\documentclass[english,12pt,a4paper]{article}
\usepackage{babel,amssymb,amsmath,graphicx,hyperref}

\begin{document}

\begin{center}
{\LARGE Comparison of the spatial properties of entrances to black holes and wormholes}\\~\\
{\large  I.D. Novikov$^{1,2,3}$, S.V. Repin$^{1}$}
\end{center}
\begin{center}
    \textit{$^1$Lebedev Physical Institute of RAS,
    AstroSpace Center, 84/32, Profsoyuznaya str., Moscow, 117997, Russia}
\end{center}
\begin{center}
    \textit{$^2$The Niels Bohr International Academy, The Niels Bohr
             Institute, \\
             Blegdamsvej 17, DK-2100, Copenhagen, Denmark}
\end{center}
\begin{center}
    \textit{$^3$National Research Center Kurchatov Institute, \\
                1, Akademika Kurchatova pl., Moscow, 123182, Russia}
\end{center}

\bigskip

\begin{abstract}
       The spatial properties of the entrances to spherically symmetric black holes and wormholes with zero and 
positive masses are compared. The properties were studied in terms of bulk and brane.  The explicit expressions 
for the entrances are found and it is shown that in both representations the spatial funnels of the entrances to 
the wormholes are the steepest for the Schwarzschild black holes, and the least steep for the wormholes with 
zero mass.

\end{abstract}

     \textbf{Keywords:} wormholes, black holes, Einstein's general theory of relativity.

     PASC number:  98.70.Ve

\section{Introduction}

    The aim of this work is to clarify the differences between the geometric spatial properties of the neighborhood 
of various relativistic objects in the framework of general relativity (GR). It should be emphasized that these 
differences do not lead directly to the differences in the visibility of these objects by an external observer. 
The fact is that the difference in visibility is associated not only with the difference in the spatial properties of these 
objects, but, even more so, with the difference in the trajectories of light beams in different gravitational fields 
around these bodies. We explore these issues in our other works. (see for example 
\cite{Zeldovich_1967, Frolov_1998, Novikov_2007, Repin_2018}).  
They are also considered in numerous publications of other authors (see for example 
\cite{Misner_1973, Thorne_2014, Shaikh_2017, Tsukamoto_2020}).   
Nevertheless, the differences in spatial properties are important for the following reasons: 1) for the analysis 
of physical processes near the entrances; 2) in relativistic theory as well as in nonrelativistic theory, the properties 
of processes depend on the geometric properties of the holes into which the flow occurs; 3) from the point of view 
of the fundamental possibility of distinguishing these objects from each other in a purely geometric way, without 
involving gravity, light and others processes. The differences obtained can be used in the future as the additional 
criteria in the observational manifestations of these objects.

     Everywhere in the paper it is supposed to be $G = 1$, $c = 1$.

\section{Objects in hyperspace (in bulk)}

    In this section, we consider the relativistic objects in a hyperspace.
     
    First we consider the spherically symmetric black holes (BH). Let us determine the shape of the surface of revolution 
in a 3-dimensional bulk, whose internal geometry in the brane (in our Universe) coincides with the 2-geometry of 
the BH equatorial ``plane''. For the definition and description of the terms bulk and brane, see, for 
example,~\cite{Thorne_2014}. As it is shown in~\cite{Landau_2012} this surface of revolution in the cylindrical 
coordinates $(r,\varphi, z)$ in the bulk is
\begin{equation}
      z_{\text{Schw}} = 2\sqrt{r_g(r - r_g)},
      \label{z_r_Schw}
\end{equation}
where $r_g = 2m$  is the gravitational radius. It corresponds to the two-dimensional metric of the equatorial section 
of the~BH. In polar coordinates in brane (in our Universe)
\begin{equation}
      dl^2 = \left(1 - \cfrac{r_g}{r} \right) dr^2 + r^2 d\varphi^2.
\end{equation}

     Find now a similar surface shape for the Ellis - Bronnikov - Morris - Thorne wormhole 
(WH)~\cite{Morris_1988, Ellis_1973, Bronnikov_1998, Bronnikov_1973}. This model of WH is mostly often used 
in theoretical astrophysics. Its metric in the two-dimensional brane is
\begin{equation}
      dl^2 = d\rho^2 + (\rho^2 + q^2) d\varphi^2,
      \label{Two_metric_MT}
\end{equation}
where $-\infty < r < \infty$, $q$~-- throat radius or in the form necessary for us
\begin{equation}
      dl^2 = \cfrac{r^2}{r^2 - q^2}\,\,dr^2 + r^2 d\varphi^2.
      \label{MT_surf_metric}
\end{equation}

      For the surface of revolution in the bulk in cylindrical coordinates, we have:
\begin{equation}
      dl^2 = dr^2 + dz^2 + r^2d\varphi^2 = 
                   dr^2 \left(1 + \left(\cfrac{dz}{dr}\right)^{\hspace{-1mm}2} \right) + r^2d\varphi^2.
      \label{MT_cyl_coords}
\end{equation}
    Comparing (\ref{MT_surf_metric}) and (\ref{MT_cyl_coords}) we obtain the differential equation
\begin{equation}
      1 + \left(\cfrac{dz}{dr}\right)^{\hspace{-1mm}2} = \cfrac{r^2}{r^2 - q^2}
\end{equation}
or
\begin{equation}
      \cfrac{dz}{dr} = \pm \sqrt{\cfrac{r^2}{r^2 - q^2} - 1}\,\,\,.
\end{equation}
Its solution is:
\begin{equation}
      z_{\text{Th}} = \pm\, q \ln\left(r + \sqrt{r^2 - q^2} \right) + z_0.
      \label{z_r_MT}
\end{equation}
Assuming $z_0 = 0$ we get an expression for the upper and lower half of the surface of revolution, each of which 
corresponds to the exit from the wormhole.

\begin{figure}[tbh]
  \centerline{
  \includegraphics[width=8cm]{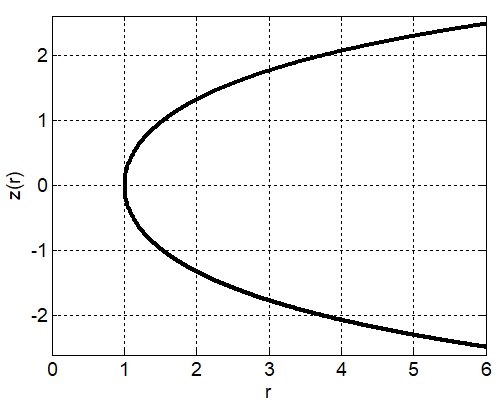}
             }
  \caption{Surface of revolution for the Morris - Thorne wormhole.}
  \label{Surface_MT}
\end{figure}

       Recall that in WH$_{\text{Th}}$ metric there are no gravitational accelerations anywhere, including the area 
outside the exits from the WH$_{\text{Th}}$. This means, particularly, that the equivalent mass of each of the inputs is 
zero, $m = 0$.

      Compare the inputs to BH and  WH$_{\text{Th}}$. They are described by the surfaces obtained by rotating 
the curves $z_{\text{Schw}}$ (\ref{z_r_Schw}) and $z_{\text{Th}}$ (\ref{z_r_MT}) around the line $z = 0$. 
Both surfaces tend to become horizontal at $r \to \infty$. However, for any finite $r$ they differ from the plane, 
and the geometry in them is different from Euclidean. The degree of difference for a given $r$ can be characterized 
by the difference $dl$ of the lengths of two close circles $r_1 = \text{const}$ and $r_2 = \text{const}$ from $2\pi dr$, 
where $r_2 = r_1 + dr$
\begin{equation}
      \cfrac{dl}{dr} = \cfrac{2\pi}{\sqrt{g_{11}}}\,\,,
\end{equation}
where $g_{11}$ is the coefficient of the corresponding metrics. For our case
\begin{equation}
      \left(g_{11}\right)_{\text{Schw}} = \left(1 - \cfrac{r_g}{r} \right)^{-1},
      \label{g11_Schw}
\end{equation}
\begin{equation}
       \left(g_{11}\right)_{\text{Th}} =  \cfrac{r^2}{r^2 - q^2}\,\,.
       \label{g11_MT}
\end{equation}
      For $ r \to \infty$ we have, respectively
\begin{equation}
      \left(g_{11}\right)_{\text{Schw}} = 1 + \cfrac{r_g}{r},
      \label{g_11_Schw_infty}
\end{equation}
\begin{equation}
       \left(g_{11}\right)_{\text{Th}} =  1 + \cfrac{q^2}{r^2}\,\,.
      \label{g_11_MT_infty}
\end{equation}

    We assume that the ``bottom'' of the spatial funnel we are investigating is $r = r_g$ for~BH and $r = q$ for~WH. 
Both curves (\ref{z_r_Schw}) and (\ref{z_r_MT})  tend to $z \to \infty$ when $r \to \infty$. 
However, we can assume our brane to be almost flat for large enough $r$, when (\ref{g_11_Schw_infty}) and
(\ref{g_11_MT_infty}) are small enough. For $r_g = q$ the determining condition for smallness is the relation:
\begin{equation}
     \cfrac{r_g}{r} \equiv \alpha \ll 1\,.
\end{equation}
It is clear that for large enough $r$ the properties of space in astrophysics will be determined by the presence of other 
objects or processes. The conclusions are qualitatively independent of the specific values of $\alpha$. Therefore, we put
\begin{equation}
     \alpha = \alpha_0 = 10^{-1}\,.
\end{equation}

\begin{figure}[tbh]
  \centerline{
  \includegraphics[width=8cm]{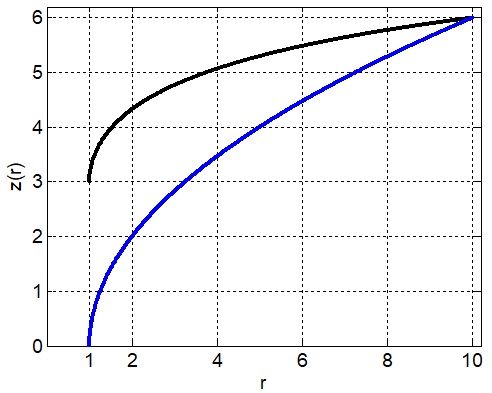}
             }
  \caption{Surface of revolution for the Morris - Thorne wormhole (above) and BH (below).}
  \label{Surface_Schw_and_MT}
\end{figure}

      Now compare the funnels in the bulk for BH and WH$_{\text{Th}}$. To do this, set $z_{\text{Schw}} = z_{\text{Th}}$ 
for $\alpha = \alpha_0$ and continue the curves for smaller~$r$ up to $r = r_g = q$.  Set the corresponding values
$z_{\text{Schw}}$, $z_{\text{Th}} = 0$. The plots are shown in Fig.~\ref{Surface_Schw_and_MT}.

      A comparison of (\ref{g11_Schw}) and (\ref{g11_MT}) for $r_g = q$ shows that the distortion of the Euclidean 
geometry when $r$  varies from large values to the inputs occurs more smoothly for WH than for BH. 
Fig.~\ref{Surface_Schw_and_MT} demonstrates that the BH funnel is ``deeper'' than the one in the case of WH.

      To avoid the misunderstandings, we recall that formally the surface of the barn in the BH case breaks off 
at~$r = r_g$, while in the case of the WH, after reaching $r = q$, the surface continues to another exit from 
the wormhole, see Fig.~\ref{Surface_MT}.

     Turn now to the case of WH with mass $m \ne 0$. Such a model was examined by Ellis \cite{Ellis_1973}; see 
also \cite{Doroshkevich_2008}.  The spatial metric of a wormhole with mass $m^*$ is: 
\begin{equation}
    dl^2 =  e^{E(\rho)}d\rho^2 + e^{E(\rho)}\left(\rho^2 + n^2 - m^2\right)
           \left(
              d\vartheta^2 + \sin^2\vartheta d\varphi^2
           \right),
            \label{Metric_Ellis}
\end{equation}
where
\begin{equation}
        E(\rho) = \cfrac{2m}{\sqrt{n^2 - m^2}} 
                        \left(\cfrac{\pi}{2} - \arctan\cfrac{\rho}{\sqrt{n^2 - m^2}} \right)
       \label{E_function}
\end{equation}
Here $n$ and $m$ are the positive constants. The $n$ value characterizes the strength of the scalar field,  and $m$ is
the effective mass $m^*$, when $\rho \to \infty$, $m \equiv m^*$. In the above example (\ref{Two_metric_MT})  
$m^* = 0$, $q = n$.

\begin{figure}[tbh]
  \centerline{
  \includegraphics[width=8cm]{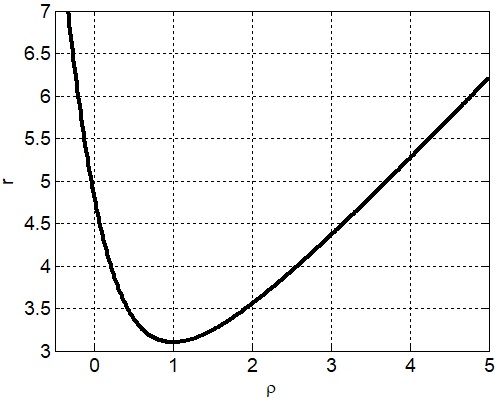}
             }
  \caption{Dependence $r(\rho)$ for the wormhole with non-zero mass.}
  \label{r_rho}
\end{figure}

     The same metric in the equatorial ``plane'' $\theta = \pi/2$ is:
\begin{equation}
    dl^2 =  e^{E(\rho)}d\rho^2 + e^{E(\rho)}\left(\rho^2 + n^2 - m^2\right)\,d\varphi^2.
    \label{Equator_WH2}
\end{equation}
      If we introduce the notation:
\begin{equation}
    r^2 =  \left(\rho^2 + n^2 - m^2\right) e^{E(\rho)},
    \label{Radius_WH2}
\end{equation}
then the function $ r(\rho)$ describes the shape of the WH$_{\text{El}}$ tunnel in the terms of $\rho$ coordinate.
The coordinate $\rho$ is often used in theoretical works, but it does not have the adequate physical meaning. The value 
of~$r$ as a function of the physical radial coordinate is given in Section~\ref{Objects_in_brane}. As an example, 
the function $ r(\rho)$ with $m = 1$, $n = \sqrt{2}$ is shown in Figure~\ref{r_rho}. It depicts the shape of 
the~WH$_{\text{El}}$ tunnel. The WH$_{\text{El}}$ entrances are located at $\rho \to \pm \infty$. In contrast to 
the case of $m = 0$, there is no symmetry here between the right and left. The exit $\rho \to \infty$ corresponds 
to $m^* > 0$, and the exit $\rho \to -\infty$ corresponds to $m^* < 0$. In this paper we will consider only the right 
branch, $m^* > 0$.

      The metric in the equatorial plane of  WH$_{\text{El}}$ can be written as:
\begin{equation}
   dl^2 = \cfrac{\rho^2 + n^2 - m^2}{\left(\rho - m\right)^2}\,\, dr^2 + r^2\, d\varphi^2,
    \label{Equator_WH2}
\end{equation}
The function $r(\rho)$ has the minimum at $\rho = m$. The value $r(m)$, i.e. the size of the throat, varies from 
$r(m) = n$ to $r(m) = en$ when $m$ runs through the values from~0 to~$n$. The throat size is always $n$ of the order 
of magnitude. We emphasize that the size of the throat is determined mainly by $n$, weakly depends on $m$ and always 
$r(m) > 2m^* \equiv r_g$.

\begin{figure}[tbh]
  \centerline{
  \includegraphics[width=8cm]{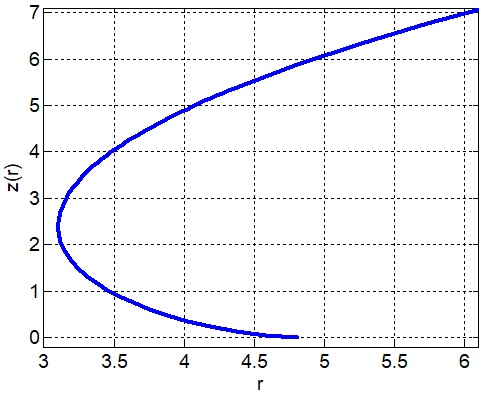}
             }
  \caption{Dependence $z(r)$ for a wormhole with a non-zero mass.}
  \label{z_func_r_E}
\end{figure}

     Imagine now the appearance of WH$_{\text{El}}$ in the bulk, as we did above for WH$_{\text{Th}}$.
Doing the same, we get for $z_{\text{El}}(r)$ a function found numerically for $m = 1$, $n = \sqrt{2}$ and shown 
in Fig.~\ref{z_func_r_E}.

      The shape is qualitatively similar to the one in Fig.~\ref{Surface_MT}, but the throat is shifted to the right and up.
In Fig.~\ref{z_func_r_E} the throat corresponds to the turning point of the curve. The branch, stretching down to 
the right, goes to the second exit. The attempts to extend the solution for $z_{\text {E}}(r)$ lead to the imaginary 
values for the integral $z(\rho)$ for $\rho < 0$ and $z < 0$. Formally, this means the inability to place the rotation 
figure in a flat three-dimensional bulk $(r, z, \varphi)$. This is not surprising, because, as we know, this second 
branch leads to the exit with a negative mass $m^* < 0$. This means that far from the output, the metric should 
correspond to the Schwarzschild solution with a negative mass. But, from the formula (\ref{z_r_Schw}) it follows 
that for $z_{\text{Schw}}$ with $r_g < 0$ and $r > 0$ the imaginary values are obtained, i.e. such a brane cannot be 
embedded in a three-dimensional flat bulk.

     Turning back to the right input with $m^* > 0$, let compare it with the entrance to the black hole. We assume that 
$r_g$ is equal to the size of the throat $r(m)$. We saw above that $r(m) > 2m^*$. The asymptotic behavior of 
deviations from the Euclidean geometry for both cases of WH$_{\text{El}}$ and BH is determined by the mass value.
Therefore
\begin{equation}
    r_g = r(m) > 2m^*\,\,.
    \label{Equator_WH2}
\end{equation}
Hence, both for WH$_{\text{El}}$ and for WH$_{\text{Th}}$, when $r$ changes from large values to the inputs, 
the Euclidean distortion increases more slowly for WHs than for BH.

\section{Relativistic objects in a brane}
\label{Objects_in_brane}

\begin{figure}[tbh]
  \centerline{
  \includegraphics[width=8cm]{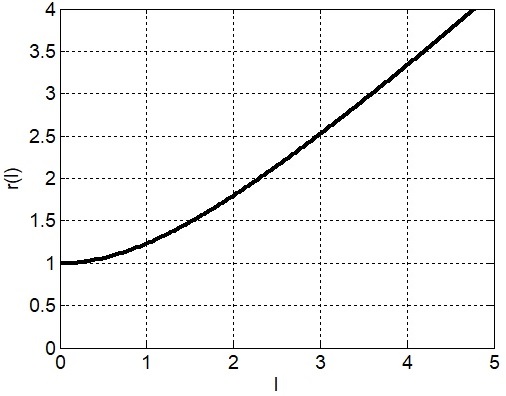}
             }
  \caption{The $r(l)$ dependence for the Schwarzschild metric. 
                The distance $l$ is counted from the gravitational radius.}
  \label{l_r_Schw_05}
\end{figure}

Let now turn to the properties of the space of holes,  when we remain in a 3-dimensional brane (in our case, in a 
2-dimensional equatorial section of the brane). We define the physical radial distance from the gravitational 
radius~$r_g$ to the point with the coordinate $r$ (in units of $r_g$) as: 
\begin{equation}
    l_{\text{Schw}} = \int\limits_1^r \sqrt{g_{11,_{\text{Schw}}}}\,dr = 
                              \sqrt{r\left(r - 1\right)} + \cfrac 12\, \ln\left(2r + 2\sqrt{r\left(r - 1\right)} - 1\right)\,.
    \label{Length_Schw}
\end{equation}
We are interested in the dependence $r_{\text{Schw}}(l_{\text{Schw}})$. This inverse function is implicitly defined 
in~ (\ref{Length_Schw}). The corresponding plot is given in Fig.~\ref{l_r_Schw_05}.

\begin{figure}[tbh]
  \centerline{
  \includegraphics[width=8cm]{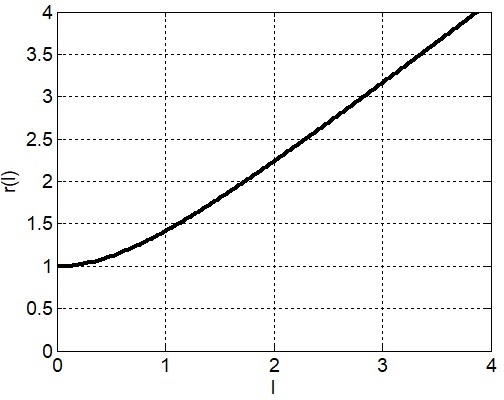}
             }
  \caption{Dependence $r(l)$ for the Morris--Thorne metric. 
                The distance $l$ is counted from the throat, where $r = q$.}
  \label{l_r_WH_MT_04}
\end{figure}

       A similar expression for the Morris--Thorne metric for the physical distance from the throat $r = q$ (in units of $ q $) is:
\begin{equation}
    l_{\text{Th}} = \sqrt{r^2 - q^2}\,.
    \label{Length_MT}
\end{equation}
The graph of the inverse function $r = r(l_{\text{Th}}) = \sqrt{\mathstrut l^2 + q^2}$ is given in 
Fig.~\ref{l_r_WH_MT_04}.

\begin{figure}[tbh]
  \centerline{
  \includegraphics[width=8cm]{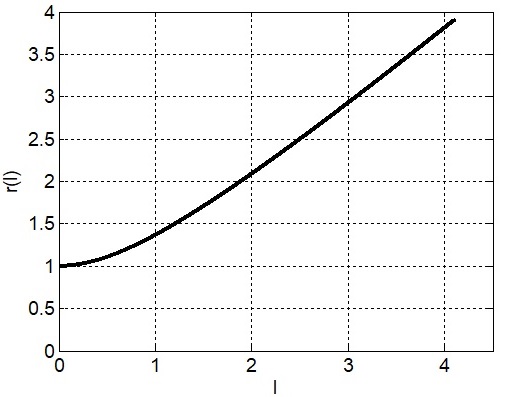}
             }
  \caption{Dependence $r(l)$ for the Ellis metric. 
                 The distance $l$ is counted from the throat, where $r = r_{min}$.}
  \label{l_r_WH_mass_plus_04}
\end{figure}

      Finally, we turn to the Ellis metric (\ref{Metric_Ellis}), (\ref{E_function}). For the exit from the WH with~$m > 0$, 
the distance from the throat $r_{min}$ to the point with the coordinate $r$ is written as:
\begin{equation}
    l_{\text{El}} = \int\limits_{r_{min}}^r \sqrt{\cfrac{\rho^2 + n^2 - m^2}{(\rho - m)^2}}\,\,dr \,.
    \label{Length_Ellis}
\end{equation}
The variable $r$ is an implicit function of $\rho$, $r_{min}$ corresponds to $\rho = m$. For our example, 
$n = \sqrt{2}$, $m = 1$. At these values $r_{min} = 3.10176639383$. We will express all distances in the units 
of $r_{min}$. The quantity $r(l)$ is shown in Fig.~\ref{l_r_WH_mass_plus_04}.

\begin{figure}[tbh]
  \centerline{
  \includegraphics[width=16cm]{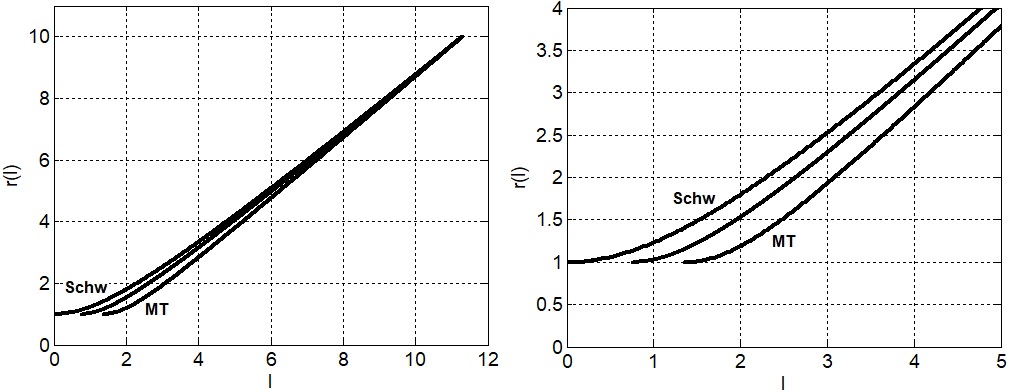}
             }
  \caption{Dependence $r(l)$ for all considered metrics. 
                 For comparison, the values of $l(r)$ are assumed to be the same for $r = 10$.}
  \label{l_r_all_011}
\end{figure}

     Fig.~\ref{l_r_all_011} shows $r(l)$ for all three cases. The values of~$r$ at~$l = 10$ are set here equal. 
The figure shows that when moving from $r = 10$ to the WH inputs the Euclidean distortion occurs most quickly for 
the Schwarzschild metric and slower for the Morris--Thorne metric.

\section{Conclusion}

      First of all, we note that the usual way to describe and depict the inputs in BH and WH as almost very similar 
is not quite correct. Both methods of visualizing the inputs in both the bulk and the brane show a significant difference 
between the metrics. The largest funnel is the Schwarzschild metric. The indicated difference should be taken into 
account when describing the processes at the entrances to the BH and WH.

\section{Acknowledgments}

       This activity has been partially supported by the KP19-270 RAS program `` Issues of the Origin and Evolution 
of the Universe Using the Methods of Ground-Based Observations and Space Research.''
       
      S.R. thanks R.E.~Beresneva, O.N.~Sumenkova and O.A.~Kosareva for the opportunity to work fruitfully 
on this problem.


\bigskip


\begin{thebibliography}{99}
\bibitem{Zeldovich_1967} Zeldovich Ya.B, Novikov I.D., \textit{Relativistic astrophysics}, 
    Moscow.: Nauka, 1967. 
\bibitem{Frolov_1998} Frolov V, Novikov I \textit{Black Hole Physics, Basic Concepts and
 New Developments} (Kluwer Academic Publishers, 1998)
\bibitem{Novikov_2007}  I.D. Novikov, N.S. Kardashev, A.A. Shatskii, \textit{Phys. Uspekhi} \textbf{50}(9), 
965 (2007).
\bibitem{Repin_2018}  S.V. Repin, D.A. Kompaneets, I.D. Novikov, V.A. Mityagina, 
   2018,  arXiv:1802.04667.
\bibitem{Misner_1973} Misner C.W., Thorne K.S., Wheeler J.A., \textit{Gravitation}, 
   San Francisco: W. H. Freeman, ISBN 978-0-393-35137-8.
\bibitem{Thorne_2014} Thorne K. \textit{The science of interstellar},
New York, London: W.W. Norton \& Co., 2014,  ISBN: 978-5-00057-536-9
\bibitem{Shaikh_2017} R. Shaikh, S. Kar, \textit{Phys. Rev. D}  \textbf{96} 044037 (2017); arXiv:1705.11008
\bibitem{Tsukamoto_2020} N. Tsukamoto, \textit{Phys. Rev. D} \textbf{101} 104021 (2020); arXiv:2004.00822v2
\bibitem{Landau_2012}  L.D. Landau, E.M. Lifshitz (1971). \textit{The Classical Theory of Fields}. 
   Vol. 2 (3rd ed.). Pergamon Press. ISBN 978-0-08-016019-1.)
\bibitem{Morris_1988} Morris M S, Thorne K S \textit{Am. J. Phys.} \textbf{56} 395 (1988)
\bibitem{Ellis_1973} Ellis H, \textit{J Math Phys} \textbf{14} 104 (1973)
\bibitem{Bronnikov_1998} K.A. Bronnikov, G. Cl{\'e}ment, C.P. Constantinidis, J.C. Fabris,
   \textit{Phys. Lett. A}, \textbf{243}, issue 3, 121 (1998)
\bibitem{Bronnikov_1973} Bronnikov K A, Acta Phys Pol \textbf{84} 251 (1973)
\bibitem{Doroshkevich_2008} A.G. Doroshkevich, N.S. Kardashev, D.I. Novikov, I.D. Novikov,
    Astronomy Reports, \textbf{52}, 616 (2008)

\end{thebibliography}
\end{document}